\documentclass[aps,prl,showpacs,notitlepage,floatfix,twocolumn,superscriptaddress]{revtex4-2}
\usepackage{amssymb}
\usepackage{amsmath}
\usepackage{amsthm}
\usepackage{amsfonts}
\usepackage{bbm}
\usepackage{graphicx}
\usepackage{braket}
\usepackage{mathtools}

\usepackage[dvipsnames]{xcolor}
\usepackage[titletoc]{appendix}

\newcommand\myshade{85}
\colorlet{myurlcolor}{MidnightBlue}

\usepackage[colorlinks,linkcolor=myurlcolor!\myshade!black,anchorcolor=myurlcolor!\myshade!black,citecolor=myurlcolor!\myshade!black,urlcolor=Magenta]{hyperref}

\usepackage{xcolor}

\definecolor{foo}{HTML}{8AAED1}
\definecolor{boo}{HTML}{F26964}
\definecolor{boo1}{HTML}{F3A7A8}
\definecolor{foo1}{HTML}{BDCFE5}
\definecolor{foo2}{HTML}{D2D3D4}

\newcommand\redcirc{{\color{boo}\bullet}\mathllap{\color{boo}\circ}}
\newcommand\bredcirc{{\color{boo1}\bullet}\mathllap{\color{boo1}\circ}}
\newcommand\bluecirc{{\color{foo}\bullet}\mathllap{\color{foo}\circ}}
\newcommand\bbluecirc{{\color{foo1}\bullet}\mathllap{\color{foo1}\circ}}
\newcommand\kink{{\color{foo2}\bullet}\mathllap{\color{foo2}\circ}}

\def\bea{\begin{equation}\begin{aligned}}
\def\eea{\end{aligned}\end{equation}}

\def\nr{n_{\redcirc}}
\def\nb{n_{\bluecirc}}
\def\nh{n_\circ}

\begin{document}

\title{Hilbert Space Fragmentation and Exact Scars of Generalized Fredkin Spin Chains}
\author{Christopher M. Langlett}
\affiliation{Department of Physics \& Astronomy, Texas A\&M University, College Station, Texas 77843, USA}
\author{Shenglong Xu}
\email{slxu@tamu.edu}
\affiliation{Department of Physics \& Astronomy, Texas A\&M University, College Station, Texas 77843, USA}

\begin{abstract}
In this work, based on the Fredkin spin chain, we introduce a family of spin-1/2 many-body Hamiltonians with a three-site interaction featuring a fragmented Hilbert space with coexisting quantum many-body scars.
The fragmentation results from an emergent kinetic constraint resembling the
conserved spin configuration in the 1D Fermi-Hubbard model in the infinite onsite repulsion limit.
To demonstrate the many-body scars, we construct an exact middle spectrum eigenstate within each fractured sub-sector displaying either logarithmic or area-law entanglement entropy.
The interplay between fragmentation and scarring leads to rich tunable non-ergodic dynamics by quenching different initial states shown through large-scale matrix product state simulations.
In addition, we provide a Floquet quantum circuit that displays non-ergodic dynamics due to sharing the same fragmentation structure and scarring as the Hamiltonian.
\end{abstract}

\maketitle

\textit{Introduction} --
In a generic closed quantum system, a simple initial state evolves unitarily in an exponentially large Hilbert space along with rapid entanglement growth.
Simultaneously, the local density matrix converges to the thermal state independent of the initial states details, a process called quantum thermalization---a direct consequence of the Eigenstate Thermalization Hypothesis(ETH)~\cite{Deutsch1991, srednicki, Rigol2008a, kim2014testing, alteth}.
ETH postulates that generic many-body systems thermalize at the individual eigenstate level, where local expectation values are the same for eigenstates with the same energy density.
In the presence of strong disorder, ETH is violated, leading to many-body localization, causing suppression of entanglement growth and retention of the initial state~\cite{Oganesyan, Nandkishore2015, Serbyn2013, Abanin2019, Schreiber2015}.

Experimental progress in various quantum platforms such as cold atoms~\cite{rubio2020floquet}, ion traps~\cite{Brydges2019,joshi2020quantum,zhu2020}, Rydberg atom arrays~\cite{bluvstein2020controlling,brow1}, and superconducting qubits~\cite{sid1,sid2} have enabled unprecedented control over probing long-time dynamics of large-scale many-body systems~\cite{google}.
This progress provides an exciting moment to study thermalization and how it is violated.
One important discovery of this renaissance is the late-time coherent oscillations in a 51-qubit Rydberg array quenched from a N\'eel state~\cite{Bernien2017}. 
Thus, providing an example of weak ETH violation~\cite{Moudgalya2018entanglement,Turner2018,lin2019exact,schecterscar}, this discovery has coalesced into an extensive study of quantum many-body scars that has unveiled a connection to emergent Lie algebras~\cite{moudgalya2020eta,onsager}, geometric frustration~\cite{Lee,Ok2019,arnab1}, lattice gauge theory~\cite{arnab2}, Floquet circuits~\cite{Pai2019,Mukherjee,Sugiura} and much more~\cite{Iadecola2019a, James2019, Zhao, Lin2020,srivatsa,wildeboer2020topological,magnifico2020}.

There has been a considerable effort to find other counterexamples to ETH.
Non-thermal behavior could arise in gauge theories~\cite{asmith1,asmith2, heyl1,karpov2020disorder, Yang2076} or in systems with combinations of conserved quantities ~\cite{Sala2020,Rakovszky2020,khemani22019,Yang2076}, where emergent kinetic constraints fracture the Hilbert space into exponentially many disconnected sub-sectors.
This phenomenon, dubbed Hilbert space fragmentation, is found in a wide range of models~\cite{tomasifrag, herviou2020many, vafek2017entanglement, lee2020frustration}, and has also been experimentally observed in the disorder-free tilted 1D Fermi-Hubbard model realized using optical lattices~\cite{schergfrag}.
As a result of the Hilbert space structure, the standard notion of thermalization does not occur.
Here, initial states thermalize within generic large sub-sectors~\cite{moudgalya2019thermalization}, with the eigenstates obeying Krylov-restricted ETH.
However, certain sub-sectors can be integrable~\cite{moudgalya2019thermalization,pozsgay2021integrable, Yang2076} or even display disorder-free localization~\cite{asmith1,asmith2, heyl1,karpov2020disorder}.
Hilbert space fragmentation provides a rich bed to study the interplay between various non-ergodic mechanisms.

What remains unexplored is whether the fragmented sub-sectors can host quantum many-body scars that weakly violate the Krylov-restricted ETH?
If yes, the combination of fragmentation and many-body scars may open a new window to richer non-equilibrium phenomena~\cite{tiltedFermi}.
This work provides a definite answer by introducing a quantum many-body model based on the deformed Fredkin spin-chain~\cite{salberger2016fredkin, salberger2017deformed, zhang2017entropy, udagawa2017finite,movassagh2016supercritical} that displays a coexistence of fragmentation and exact many-body scars.
We show that by embedding the Fredkin chain's ground state~\cite{Shiraishi2017,Shiraishi2019} into the middle of the spectrum, the Hilbert space splits into exponentially many disconnected sub-sectors.
We find an exact ETH violating eigenstate within \textit{every} sub-sector, while numerics reveal further scarring.
Due to the fragmentation and many-body scar states, the Hamiltonian displays rich tunable non-ergodic dynamics presented through large-scale tensor network simulations.
Interestingly, the non-ergodic dynamics are closely related to domain-wall melting and propagation in the XXZ spin chain with a boundary magnetic field.
Furthermore, we substantiate our results by designing a Floquet quantum circuit that hosts the same dynamical properties.

\textit{Model} --
The simplest version of the model we introduce is the dressed Heisenberg chain,
\bea\label{eq:dressed_heisenberg} 
H = \sum\limits_{i=2}^{N-2} \sigma^{z}_{i-1} \left(1 - \vec{\sigma}_i \cdot \vec{\sigma}_{i+1} \right) + \left(1-\vec{\sigma}_{i}\cdot \vec{\sigma}_{i+1} \right) \sigma^{z}_{i+2}
\eea
The Hamiltonian above conserves total magnetization, $S_{z} = \sum_i \sigma^z_i$,
and displays a symmetric spectrum around zero due to anti-commuting with the operator, $C = \prod_{i} \sigma^{x}_{i}$~\cite{Schecter2018}.
The off-diagonal elements of the Hamiltonian are given by the following spin moves,
\begin{equation}\label{eq:move}
    \ket{\uparrow\uparrow\downarrow} \longleftrightarrow     \ket{\uparrow\downarrow\uparrow}, \quad \ket{\downarrow\uparrow\downarrow} \longleftrightarrow -     \ket{\uparrow\downarrow\downarrow}.
\end{equation}
The above spin-moves are analogous to the controlled-swap or Fredkin quantum gate.
From destructive interference between the spin moves there is also $U(1)$ conservation of domain-wall number, $n_{dw} = \sum_i \sigma^z_i \sigma^z_{i+1}$, which with $S_z$ leads to the Hilbert space fragmentation~\cite{Yang2076}.
For example, the N\'eel state and the domain-wall state, $\ket{\uparrow \cdots \uparrow \downarrow \cdots \downarrow }$, are both eigenstates corresponding to frozen sub-sectors.

The above model is generalized to a family of frustration-free Hamiltonians parameterized by $q$
\begin{equation}\label{eq:embed_fredkin}
 H(q) = \frac{4}{q} \sum\limits_{i=2}^{N-2}\left( P_{i-1}^{\uparrow}\ket{\Phi}\bra{\Phi}_{i,i+1} -\ket{\Phi}\bra{\Phi}_{i,i+1}P_{i+2}^{\downarrow}\right),
\end{equation}
where the state $\ket{\Phi}_{i,i+1}=\ket{\uparrow\downarrow}-q\ket{\downarrow\uparrow}$ is on site $i$ and $i+1$ and $P^{\uparrow(\downarrow)}$ are the spin-$1/2$ projection operators.
At $q=1$, the Hamiltonian recovers Eq.~(\ref{eq:dressed_heisenberg}).
When both term in the Hamiltonian are positive, the Hamiltonian becomes the deformed Fredkin chain~\cite{salberger2016fredkin, salberger2017deformed}, which is known for solvable ground states with unique entanglement properties. 
The minus sign in Eq.~(\ref{eq:embed_fredkin}) introduces destructive interference preventing spin configurations from connected and therefore fractures the Hilbert space into invariant sub-spaces.
More importantly, we construct an exact eigenstate with sub-volume law entanglement entropy for \textit{each} sub-sector which due to the frustration free form of Eq.~(\ref{eq:embed_fredkin}) is valid for arbitrary $q$.
As a result, the model described in Eq.~(\ref{eq:embed_fredkin}) exhibits fragmentation coexisting with quantum many-body scars that lead to a rich class of non-ergodic dynamics. 

\textit{Fragmentation and Effective Hamiltonian} --
 \begin{figure}
    \includegraphics[width=\columnwidth]{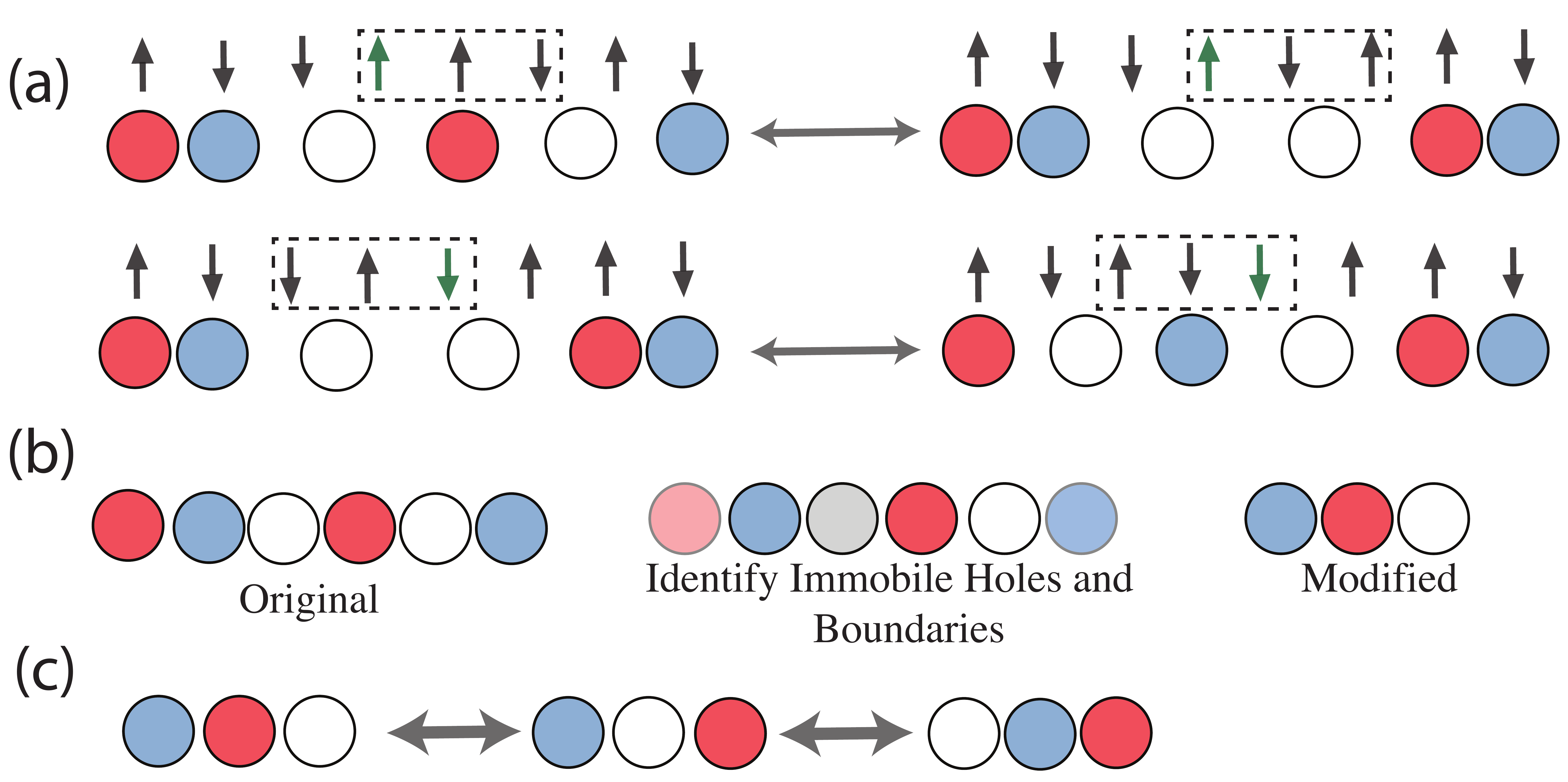}
    \caption{
    (a) Visualization of the controlled-swap spin flip in Eq.~\ref{eq:move} on original spins and the effective particle picture. The control bit is highlighted in green. The spin flip corresponds to hole hopping through the particles.
    (b)Immobile holes, $(\kink)$, and left(right) boundary particles denoted, $\bredcirc(\bbluecirc)$, are inert and removed leaving a modified sequence.
    (c) The spin flips correspond to nearest-neighbor hopping of the holes in the modified sequence.}
    \label{fig:hsf}
\end{figure}
We characterize the fragmentation structure by developing an effective description of Eq.~(\ref{eq:embed_fredkin}) in analogy with the 1D Fermi-Hubbard model at infinite-$U$, where the electron spin configuration is conserved.
Given a spin configuration, we first map the two-site state, $\ket{\downarrow \uparrow} \rightarrow \ket{\circ}$, and then map the remaining individual spins to colored particles, $\ket{\uparrow (\downarrow)}\rightarrow\ket{\redcirc(\bluecirc)}$.
Consequently, the spin moves, Eq.~(\ref{eq:move}), produce a kinetic constraint forbidding red(blue) particle exchange.
Analogous to the electrons spin configuration of the infinite-$U$ Hubbard model, the color order of the particles is a Hamiltonian invariant.
The fractured sub-spaces are constructed by all spin states corresponding to the same color sequence.
While maintaining this constraint, the spin flips, Eq.~(\ref{eq:move}), correspond to particle hopping, shown in Fig.~\ref{fig:hsf}(a).
There is a further constraint that at least one hole sits between a left blue particle and a right red particle.
The configuration in the particle picture has the form, $\bluecirc\circ\redcirc$, or $\downarrow\downarrow\uparrow\uparrow$ in the spin representation, which is frozen as a result of conservation of $n_{dw}$ and $S_z$.
The same situation occurs for the holes between the left(right) boundary and a red(blue) domain (see SM~\cite{SupMat} and references therein
~\cite{kim2018efficient, ono2017implementation, kang2020optical, gambetta2020long, feng2020quantum, petiziol2020quantum}
).
In addition, the red(blue) particle on the left(right) boundary are immobile as well because the Hamiltonian conserves the boundary spins.
Removing these immobile degrees of freedom (Fig.~\ref{fig:hsf}(b)), the action of spin flips Eq.~(\ref{eq:move}) are understood as nearest-neighbor hopping of the holes through colored particles with fixed color order shown in Fig.~\ref{fig:hsf}(c).
We denote the number of mobile red(blue) particles and holes as $\nr(\nb)$ and $\nh$.
The dimension of each fractured sub-sector is simply the binomial number,
$\dim = \binom{\nr + \nb + \nh}{\nh}$.
The largest sub-sector occurs when $\gamma=\nh/N =(5-\sqrt{5})/10$ with $\dim\sim ((1+\sqrt{5})/2)^N$ which is exponentially smaller than the total Hilbert space $2^N$, while single dimension sub-sectors(\textit{frozen}) arise when either $\nr+\nb=0$ or $\nh=0$.

We further construct an effective Hamiltonian equivalent to Eq.~(\ref{eq:embed_fredkin}) that directly acts on the particles 
\bea\label{eq:effective}
H(q) = & -t \sum \limits _{i} \left(c^\dagger_{i,\redcirc} c^{}_{i+1, \redcirc} -c^\dagger_{i,\bluecirc} c^{}_{i+1, \bluecirc} \right) +h.c.\\
& +V(q) \sum\limits_{i} (n_{i,\redcirc}-n_{i+1,\redcirc})^2 - (n_{i,\bluecirc}-n_{i+1,\bluecirc})^2 \\
& +h(q) \sum\limits_{i} (n_{i,\redcirc}-n_{i+1,\redcirc}) + (n_{i,\bluecirc}-n_{i+1,\bluecirc}).
\eea
where $c_{i,\redcirc(\bluecirc)}$ is the annihilation operator of a red(blue) particle at site $i$, and $n_{i,\redcirc(\bluecirc)}$ counts the onsite color particle number.
The on-site double occupied states are projected out from the Hilbert space, leading to the conserved color order of the particles that labels each fractured sub-sector.
Here the hopping parameter, $t = 4$, nearest-neighbor interaction, $V(q)=2(1/q+q)$, and onsite potential, $h(q)=2(1/q-q)$.
The onsite potential cancels in the bulk leaving only a boundary term, $H_{\partial}=h(q)(n_{1,\redcirc} + n_{1,\bluecirc} 
-n_{L,\redcirc}-n_{L,\bluecirc})$.
Through a Jordan-Wigner transformation, Eq.~(\ref{eq:effective}) is equivalent to an XXZ ladder with a ferromagnetic and anti-ferromagnetic coupling on each leg with the constraint that the two spins on each rung cannot be up simultaneously.

\begin{figure}
\includegraphics[width=\columnwidth]{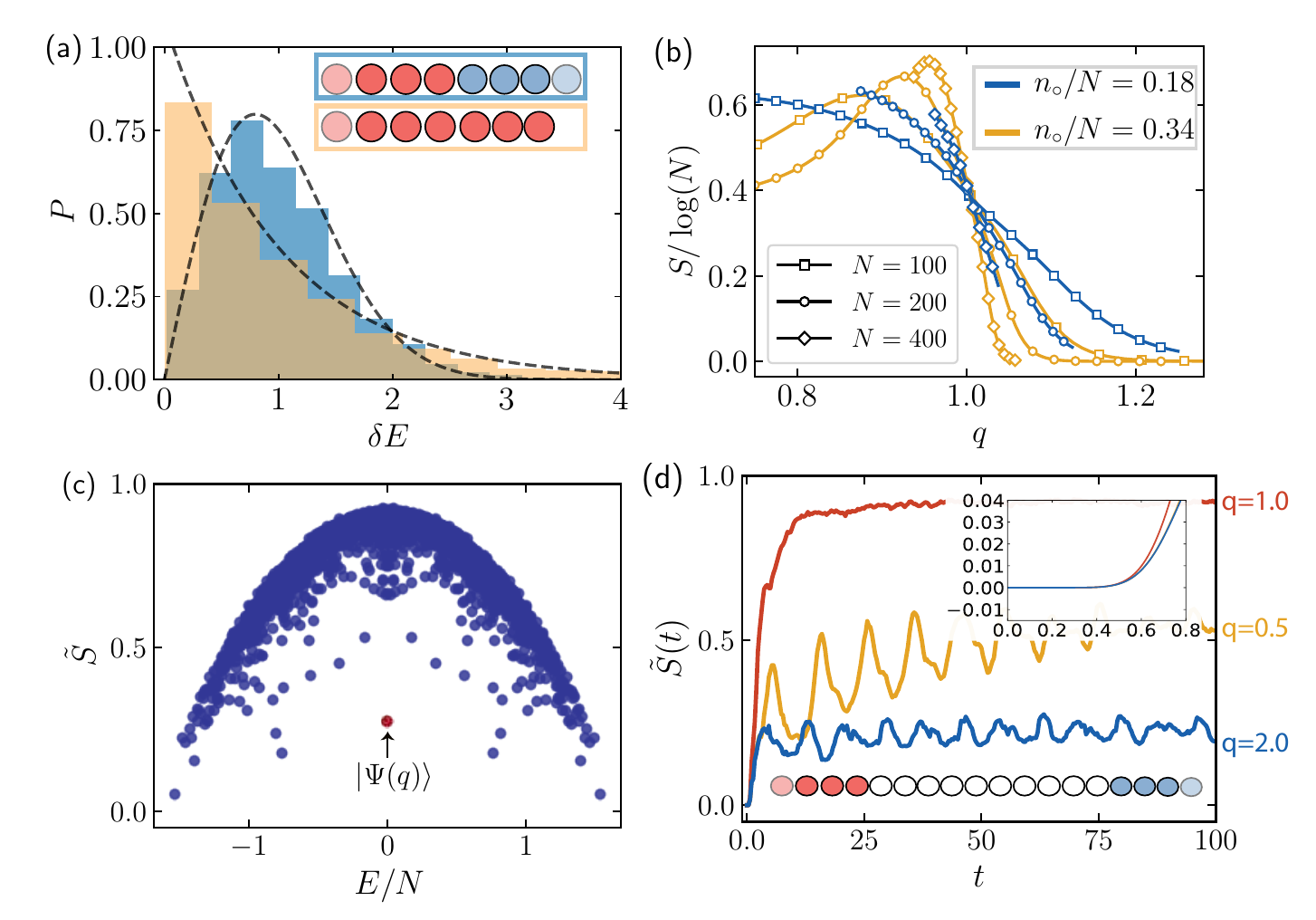}
\caption{(a) At $q=1$, two fractured sub-sectors display Wigner-Dyson and Poisson Many-body level statistics. Each sub-sector contains 10 holes with $\dim=8008$ and are labeled by the color sequence sketched in the figure. 
(b) The scaling of the R\'enyi entropy of the exact eigenstate from the MPS construction for the chaotic sub-sector. The ratio is fixed as $N$ increases from 100 to 400. The entanglement entropy scales as $\log (N)$ at $q=1.0$, but decays to zero for $q>1$. For $q<1$, the scaling depends on the ratio $\nh/N$. 
(c) Half-chain normalized R\'enyi entropy, $\tilde{S}$, of each eigenstate in the chaotic sub-sector. The low lying states, including the exact eigenstate in Eq.~(\ref{eq:scar}) highlighted in red, indicate quantum many-body scars. 
(d) Dynamics of $\tilde{S}$ for the initial state within the chaotic sub-sector using exact diagonalization. Here $N=28$ with $\nh=10$. For early times, $\tilde{S}$ stays zero until $t\sim 0.6$ shown in the inset. The late time behavior depends on $q$.}
\label{fig:2}
\end{figure}

\textit{Thermalization and Integrability} --
The color order is central to the thermalization properties of the sub-sectors. 
When the sequence is monochromatic, the effective Hamiltonian is equivalent to the integrable XXZ chain with an easy-axis anisotropy parameter, $\Delta = V(q)/4 \geq 1$. 
On the other hand, a generic color sequence is expected to display Krylov-restricted thermalization~\cite{moudgalya2019thermalization}.
As an example of the different behaviors we consider two sub-sectors at $q=1.0$(Fig.~\ref{fig:2}(a) inset) with the same dimension, and study the level statistics quantified with the average ratio parameter, $r_{n}=\min(\delta E_{n},\delta E_{n+1})/\max(\delta E_{n},\delta E_{n+1})$~\cite{MBLPal,Atas2013,Oganesyan}, where $\delta E_{n} =E_{n+1}-E_{n}$ is the gap between subsequent energies.
We find for the single color, $\langle r \rangle \approx 0.386$, as expected for an integrable system.
While for the sub-sector characterized by, $\bredcirc\redcirc\cdots\redcirc\bluecirc\cdots\bluecirc\bbluecirc$, $\langle r \rangle \approx 0.529$ which is consistent with the Wigner-Dyson result, indicating chaos.
This sub-sector remains Wigner-Dyson for $0.5\leq q\leq2.0$ (see Fig.~S1~\cite{SupMat}).

\textit{Exact many-body scars} --
The system's most remarkable feature is that each fractured sub-sector hosts one exact eigenstate for arbitrary $q$.
\bea
\label{eq:scar} 
\ket{\Psi(q)}=\frac{1}{\mathcal{N}(q)}\sum_{i}^{\text{dim}} e^{-\log(q) \hat{\mathcal{P}}/2} \ket{\psi}_{i},
\eea
where $\hat{\mathcal{P}} = \sum_i i \sigma^z_i$ is the dipole operator in the spin representation,  the summation is over all states in the computational basis in each fractured sub-sector, and the normalization factor, $\mathcal{N}(q)^{2} = \text{tr}(q^{-\hat{\mathcal{P}}})$. 
The eigenstate in each fractured sub-sector generalizes a Rokhsar-Kivelson state to the middle of the spectrum~\cite{RK1, RK2,RK3}.
To verify Eq.~(\ref{eq:scar}) is an eigenstate of the Hamiltonian Eq.~(\ref{eq:embed_fredkin}), we construct a product state $\exp(-\log(q)\hat{\mathcal{P}}/2)\prod_{i=1}^N\ket{\uparrow+\downarrow}_i = \sum \mathcal{N}(q)\ket{\Psi(q)}$,  a superposition of $\ket{\Psi(q)}$ from each disconnected subsector. One can verify that this product state is annihilated by the Hamiltonian, i.e., a zero energy eigenstate. As a result, $\ket{\Psi(q)}$ from each disconnected subsector must also be a zero-energy eigenstate. 
Due to the wavefunction structure and hole conservation, the half-chain entanglement $S$ of Eq.~(\ref{eq:scar}) is upper bounded by $\log(\min(2\nh+1, N-2\nh+1))$.
This result applies for all R\'enyi entropies, $S^{(\alpha)}=\log\text{tr}(\rho^\alpha)/(1-\alpha)$.
Throughout this work we utilize the second-order R\'enyi$(\alpha=2)$, denoted as $S$.
This is substantiated by explicit construction of a matrix product state of Eq.~(\ref{eq:scar}) shown in SM~\cite{SupMat}.
Consequently, $S$ scales at most logarithmically with $N$, in direct violation of ETH, which postulates a volume-law entanglement scaling for middle spectrum eigenstates.
Therefore, $\ket{\Psi(q)}$ is scarred within ergodic sub-sectors such as the one associated with, $\bredcirc\redcirc\cdots\redcirc \bluecirc\cdots\bluecirc \bbluecirc$~\footnote{If the sequence only contains red(blue) particles, the effective Hamiltonian is positive(negative).
Therefore $\ket{\Psi(q)}$ is the ground (ceiling) state in this case.}.

We analyze the entanglement scaling of $\ket{\Psi(q)}$ as a function of $q$, $\nr+\nb$ and $\nh$ see SM~\cite{SupMat}.
When $\nr+\nb$ or $\nh$ stay constant with $N\rightarrow \infty$, the entanglement has area law scaling for any $q$ due to the upper bound.
Utilizing a large $N$ analysis in the sub-sector, $\bredcirc\redcirc\cdots\redcirc \bluecirc\cdots\bluecirc\bbluecirc$, we show that when $\nr + \nb$ and $\nh$ scale linearly with system size, $S \sim \log (N)$ at $q=1$.
As $q$ increases, $S$ rapidly approaches zero(see Fig.~\ref{fig:2}(b)) because $\ket{\Psi(q)}$ is limiting towards a product state with the largest dipole amplitude.
When $q<1$, the eigenstate is approximately a superposition of $(\nh+1)$ states with the smallest dipole moment amplitude.
Depending on whether $\nh/N>1/4$, the state exhibits area or logarithmic entanglement. 
This analysis is confirmed by direct calculation of $S$ using MPS for large system size up to $N=400$, illustrated in Fig.~\ref{fig:2}(b).
We further study the normalized eigenstate entanglement entropy, $\tilde{S}=S/\log(\sqrt{\text{dim}})$, using exact diagonalization for $N=28$, shown in Fig.~\ref{fig:2}(c).
Intriguingly, the plot clearly demonstrates outlying states significantly below the ETH-like curve, indicating multiple many-body scarred states distinct from Eq.~(\ref{eq:scar}). 

\textit{Dynamics} --
These scarred states manifest themselves in the quantum dynamics of special initial states. 
We identify the product state,  $\ket{\psi}=\ket{\bredcirc\redcirc\cdots\redcirc\circ\cdots\circ\bluecirc\cdots\bluecirc\bbluecirc}$, in the effective picture with the largest dipole moment amplitude due to the high overlap with the exact eigenstate for $q>1.0$.
This state has zero-energy density and by ETH should be thermal.
We study the half-chain entanglement dynamics of $\ket{\psi}$ for various $q$. 
A first indication of atypical dynamics is a plateau of zero-entanglement for $t<0.6$, present for each $q$ and absent in otherwise random states, inset Fig.~\ref{fig:2}(d).
This follows from the time-scale of the rightmost red particle and leftmost blue particle to propagate to the center and scatter.
When $q=1.0$, $\tilde{S}$, quickly saturates, while at $q=2.0$, the entanglement remains finite at $\tilde{S}\sim 0.2$, far below the thermal expectation due to being approximately an eigenstate.
Late-time coherent oscillations arise at $q=0.5$ and are absent at $q=2.0$ despite the effective Hamiltonian differing only by $H_{\partial}$. 

\begin{figure}
\includegraphics[width=\columnwidth]{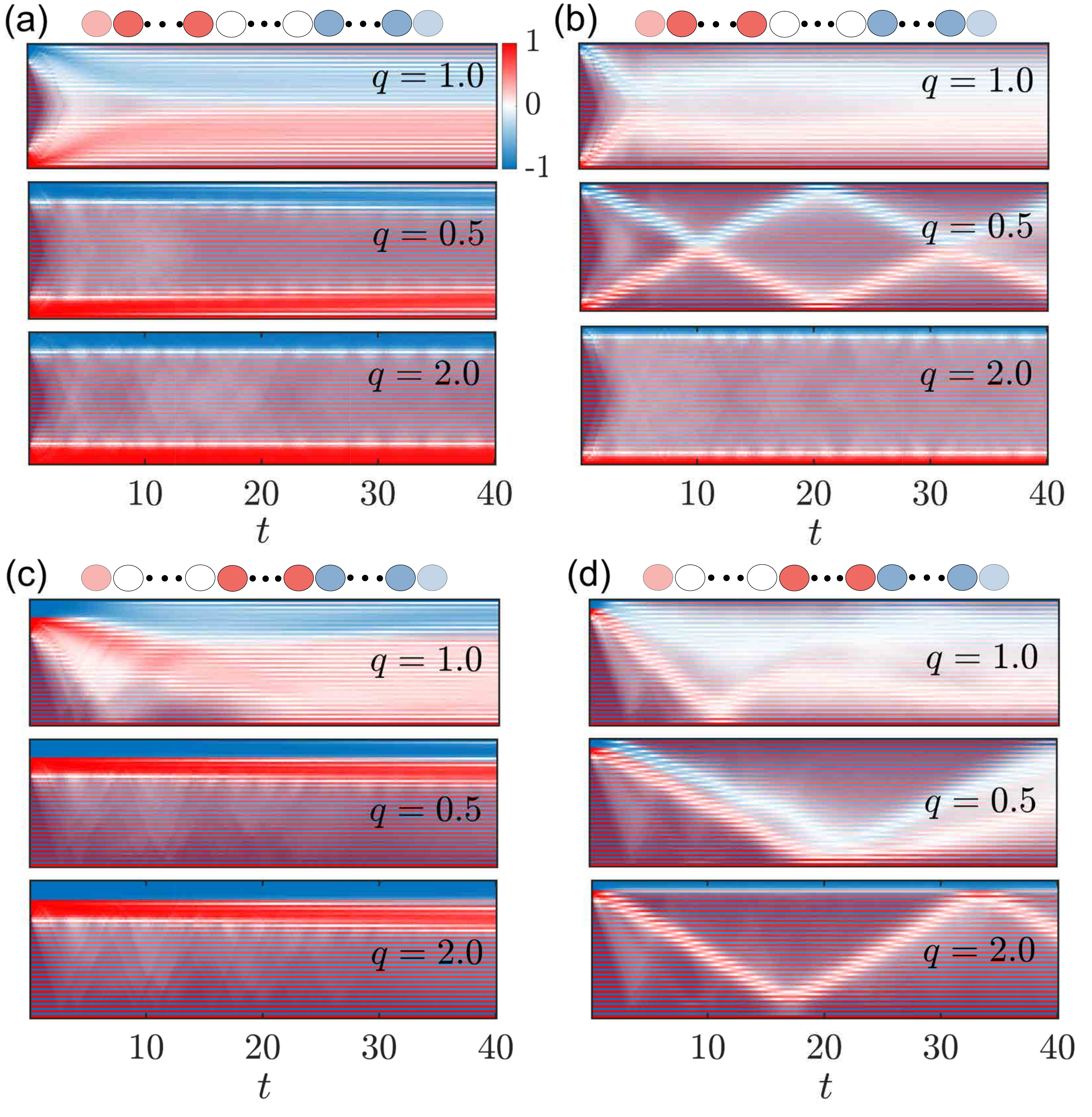}
\caption{Dynamics of $\braket{\sigma^z_i(t)}$ at $q=0.5, 1.0, 2.0$ for two domain-wall initial states for $N=56$. We consider two different domain lengths. In (a) and (c), $\nr=\nb=7$ and $\nh=20$. In (b) and (d), $\nr=\nb=3$ and $\nh=24$.
We simulate the dynamics using TDVP-MPS with bond dimension 48 and time step $dt=0.04$ for $t=40$.
Technical details of TDVP and result from exact diagonalization can be found in SM~\cite{SupMat}.
}
\label{fig:3}
\end{figure}

The various non-ergodic behavior arise from domain-wall dynamics in the effective Hamiltonian Eq.~(\ref{eq:effective}).
The state, $\ket{\psi}$, contains two domain-walls between the red(blue) particle domain and the hole domain.
In the spin representation the domain walls are between fully polarized and N\'eel configurations.
During unitary evolution, the domain-walls melt independently before meeting at the center, Fig.~\ref{fig:3}(a).
The effective Hamiltonian structure suggests the domain-wall melting is related to the XXZ chain for $\Delta \geq1$ with an additional boundary term.
For the XXZ chain, the domain-wall state is near the ground state, where the melting is super-diffusive at $\Delta=1$ but nearly static at $\Delta>1$~\cite{ljubotina2017class,mossel2010relaxation}.

The non-ergodic domain-wall dynamics prepared in $\ket{\psi}$ is simulated using large-scale tensor network simulations based on the time-dependent variational principle(TDVP)~\cite{Haegeman,cirac}.
We first consider a chain of length $N=56$ with $\nr=\nb=7$ and $\nh=20$, with small system exact diagonalization found in SM~\cite{SupMat}. 
Fig.~\ref{fig:3} plots the space-time dynamics of $\braket{\sigma^z_i(t)}$ in the spin representation which displays domain-wall melting at $q=1.0$ but static at $q=2.0$, similar to the XXZ chain.
While at $q=0.5$ the domain-walls do not melt, the two domains do however slowly leave the boundary unlike at $q=2.0$.
This difference becomes more prominent for shorter domains, shown in Fig.~\ref{fig:3}(b).
We emphasize that the bulk terms in the effective Hamiltonian are the same for $q=0.5$ and $q=2.0$, with the different dynamics arising due to the boundary term, $H_\partial$.
The boundary term changes the energy cost of moving the domain from $V(q)=2(q+1/q)$ to $V(q)-h(q)=4q$, a significant reduction for $q<1$.
The consequence is ballistic domain propagation illustrated in Fig.~\ref{fig:3}(b) for $q=0.5$.
In contrast, for $q>1$ the boundary term increases the energy cost, rendering the domain static.
This also explains the entanglement growth in Fig.~\ref{fig:2}(d) between $q=0.5$ and $q=2.0$, where the coherent oscillations for $q=0.5$ are attributed to domain scattering.
This mechanism also suggests non-ergodic dynamics occur for initial states with the red(blue) domains of varying length at different positions.
As an example, we study the quench dynamics of another initial state, $\ket{\bredcirc\circ\cdots\circ\redcirc\cdots\redcirc \bluecirc \cdots \bluecirc\bbluecirc}$, with zero-energy density, shown in Fig.~\ref{fig:3}(c) and (d).
The nearest-neighbor energy cancels in the configuration, $\cdots\redcirc\circ\bluecirc\cdots$, which permits collective hopping of the red domain without changing the configuration energy. 
As a result, the red domain always displays ballistic propagation, while at $q=0.5$ the blue domain moves from an energy reduction due to $H_{\partial}$, similar to $\ket{\psi}$ at $q=0.5$.

\begin{figure}[t]
\includegraphics[width=1\columnwidth]{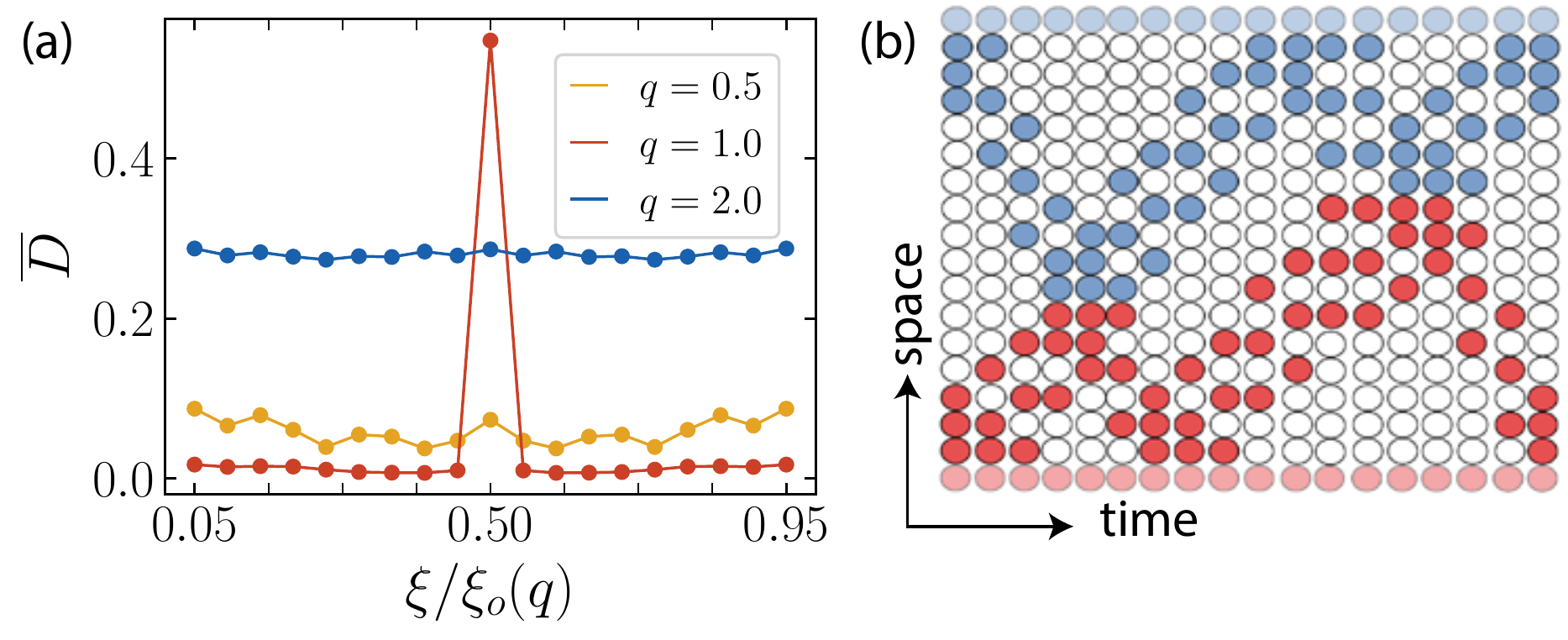}
\caption{
(a) Thermal deviation for the Floquet circuit within the sub-sector $N=28$ with $\nr=\nb=3$ and $\nh=10$ prepared in the state used in Fig.~\ref{fig:2}(d). The result is averaged over late-time driving periods between 3200 and 3600.
For $q=1.0$ there is a sudden jump at $\xi/\xi_{o}(q) = 0.5$ corresponding to the CA.
(b) CA dynamics for one period of the classical trajectory.}%
\label{fig:circuit_dynamics}
\end{figure}

\textit{Quantum Circuit} --
We extend the Hamiltonian in Eq.~(\ref{eq:embed_fredkin}) to a unitary Floquet quantum circuit sharing the same properties.
The circuit contains three-site unitary gates which has immediate implementation~\cite{lanyon2011universal,barends2015digital,britton2012engineered,zhang2020highfidelity}. 
The Floquet operator takes the form
\bea\label{eq:circuit}
&U_{T}(\xi,q)= \prod\limits_{i=1}\prod^{3}_{l=1} U^\uparrow_{3i+l-3} U^\downarrow_{3i+l}\\
&U_i^\uparrow= e^{\frac{-4i\xi}{q}\left(P_{i}^{\uparrow} \ket{\Phi}\bra{\Phi}_{i+1,i+2}\right)}, \ \ 
U_{i}^\downarrow= e^{\frac{+4i\xi}{q}\left(P_{i}^{\downarrow} \ket{\Phi}\bra{\Phi}_{i-2,i-1}\right)}.
\eea
The parameter, $\xi$ is in the interval $[0, \xi_{o}]$ with $\xi_{o}(q)=q\pi/(2+2q^2)$ since $U_T(\xi+\xi_o,q)=U_T(\xi,q)$.
The circuit configuration and gate decomposition are shown in SM~\cite{SupMat}.
When $\xi\rightarrow 0$ the circuit is the Trotterization of the Hamiltonian dynamics.
For general $\xi$, the circuit exhibits the same fractured Hilbert space as the Hamiltonian and importantly the scar state in Eq.~(\ref{eq:scar}) remains an eigenstate of the Floquet operator. Typically states thermalize to infinite temperature; however, here, similar non-ergodic dynamics are observed.
In addition, at the point $(\xi=\pi/8, q=1.0)$, Eq.~(\ref{eq:circuit}), is a classical cellular automaton(CA) which has recently received a reviving interest in non-equilibrium dynamics~\cite{Wilkinson2020exact, iadecola2020nonergodic, Gopalakrishnan2018}.
Here the update rules are given by the Fredkin gate
\bea\label{eq:fredkin}
U^{\uparrow}_{i}&=e^{-i\frac{\pi}{4} P_i^\uparrow (1-\vec{\sigma}_{i+1}\cdot\vec{\sigma}_{i+2}) }, \ \
U^{\downarrow}_{i}=e^{+i\frac{\pi}{4} P_i^\downarrow (1-\vec{\sigma}_{i-1}\cdot\vec{\sigma}_{i-2}) }.
\eea
We quantify the thermal deviation of the dynamics starting from the same initial state in Fig.~\ref{fig:2}(d) with,
$D(t)=\sum_{i}\left|\langle \sigma_{i}^{z}(t) \rangle -\langle \sigma_{i}^{z}\rangle_{\text{th}}\right|/N$, which we then average over late-time driving periods. 
The result, $\overline{D}$, for different $\xi$ and $q$ is shown in Fig.~\ref{fig:circuit_dynamics}(a), demonstrating that driving maintains the non-thermal properties.
The large peak at $q=1.0$ and $\xi/\xi_{o}(q)=0.5$ is attributed to the classical point, showing the maximal deviation from thermalization. The corresponding classical dynamics in the effective picture is plotted in Fig.~\ref{fig:circuit_dynamics}(b).  

\textit{Conclusion} --
In this work, we introduced a spin-1/2 quantum Hamiltonian with three-site interactions that exhibit a fractured Hilbert space hosting chaotic and integrable sub-sectors.
We construct an exact eigenstate with sub-volume law entanglement, which violates Krylov-restricted ETH in chaotic sub-sectors.
Quenching from domain-wall initial states in chaotic sub-sectors leads to rich non-ergodic domain-wall dynamics, whereas typical states thermalize.
In connection with recent advancements in noisy intermediate quantum devices~\cite{Preskill2018quantumcomputingin}, we also provide a quantum circuit sharing the Hamiltonian's fundamental dynamical properties.
From the union of fragmentation and many-body scars, our system enables control over various non-equilibrium phenomena by changing the initial states.
This work leads to many interesting future directions, such as studying quenched disorder,  quasi-periodic color sequences, and extending the construction to the colored Fredkin model.

\textit{Acknowledgement} -- S.X and C.M.L thank 
Lakshya Agarwal, Zhi-Cheng Yang and Brian Swingle for helpful comments on the manuscript. The numerical simulation in this work was conducted with the advanced computing resources provided by Texas A\&M High Performance Research Computing.
\bibliography{references}
\end{document}


\title{Supplementary Material: Hilbert Space Fragmentation and Exact Scars of Generalized Fredkin Spin Chains}
\author{Christopher M. Langlett}
\affiliation{Department of Physics \& Astronomy, Texas A\&M University, College Station,Texas 77843, USA}
\author{Shenglong Xu}
\affiliation{Department of Physics \& Astronomy, Texas A\&M University, College Station,Texas 77843, USA}

\begin{abstract}
In this Supplementary Material, we provide detailed information regarding the map from spins to particles, an analysis of the exact eigenstate's entanglement scaling, simulation of the Hamiltonian using matrix product states that incorporate the fractured Hilbert space, as well as more data on the dynamics using exact diagonalization.
We explicitly demonstrate the boundary's role by studying the effective model with and without the boundary term using exact diagonalization. 
We also provide efficient gate decomposition of the Floquet circuit discussed in the main text and show the circuit's non-ergodic dynamics for different parameter space points.
\end{abstract}
\maketitle
\tableofcontents

\section{Controlled-SWAP Spin Moves and Mapping Spins to Particles}
\label{appsec:map}
In this section we first discuss how the destructive interference arises and then outline how to develop the color ordered configurations.
Recall that the spin moves of the Hamiltonian, Eq.(2) perform a controlled-SWAP operation on three-spin sequence as follows, $\ket{\uparrow\uparrow\downarrow} \longleftrightarrow     \ket{\uparrow\downarrow\uparrow}$ and $\ket{\downarrow\uparrow\downarrow} \longleftrightarrow -\ket{\uparrow\downarrow\downarrow}$.
Due to the destructive interference of the two spin moves, the allowed spin moves on four-spins are
\bea
\ket{\uparrow\uparrow\downarrow\uparrow} \longleftrightarrow     \ket{\uparrow\downarrow\uparrow\uparrow}\quad
\ket{\downarrow\downarrow\uparrow\downarrow} \longleftrightarrow -\ket{\downarrow\uparrow\downarrow\downarrow}
\eea
As a result, the domain-wall and N\'eel states are eigenstates of the Hamiltonian.
As a demonstration consider the domain-wall state, $\ket{\uparrow\uparrow\cdots\uparrow\uparrow\downarrow\downarrow\cdots\downarrow\downarrow}$.
According to the map, this state becomes: $\ket{\uparrow\uparrow\cdots\uparrow\downarrow\uparrow\downarrow\cdots\downarrow\downarrow}-\ket{\uparrow\uparrow\cdots\uparrow\downarrow\uparrow\downarrow\cdots\downarrow\downarrow}$, which clearly cancels.
This interfering behavior prevents sub-sectors from being connected that otherwise are in the standard Fredkin model.

Due to the spin moves' destructive interference, each sub-sector carries a domain-wall number accompanied by a $U(1)$ charge, which becomes manifest through an effective particle-hole mapping.
The sub-sectors are characterized by a color string of particles with an imposed kinetic constraint of disallowed particle exchange.
The set of all spin states that map to the same color sequence constructs a fragmented sub-sector of the Hilbert space.
In order to determine these sequences from a given state, we apply the steps discussed below.

Given a spin state in the full Hilbert space first we map every sequence, $(\downarrow\uparrow)\longrightarrow\circ$, denoted as a hole.
Then every individual remaining spin becomes a color particle, $\downarrow(\uparrow)\longrightarrow\bluecirc(\redcirc)$.
The unique label for the sub-sector is found by removing all the holes, leaving a colored string where the order of particle color must be maintained.
The spin moves in this picture correspond now to a particles hopping through a hole background.
However, there are certain sequences when an immobility arises.
These possibilities are summarized as, (i) $\cdots\bluecirc\circ\redcirc\cdots$, (ii) $|\circ\redcirc\cdots$, (iii) $\cdots\bluecirc\circ |$, where $|$ stands for the boundary. The holes in the above configurations are immobile due to conservation of $n_{dw}$ and $S_z$. 
More generally, if there are multiple holes in the above configurations, a single immobile hole will be left after hopping.
We demonstrate this with the following example.

Consider the state, $\downarrow\downarrow\downarrow\uparrow\downarrow\uparrow\downarrow\uparrow\uparrow\uparrow$.
First, we replace each two-spin sequence, $(\downarrow\uparrow)$, with, $\circ$.
Our state now becomes, $\downarrow\downarrow\circ\circ\circ\uparrow\uparrow$.
Now map the remaining individual spins to colored particles resulting in the state, $\bluecirc\bluecirc\circ\circ\circ\redcirc\redcirc$.
Consider the sequence of the blue particle hopping to the right,$\bluecirc\bluecirc\circ\circ\circ\redcirc\redcirc\longrightarrow\bluecirc\circ\bluecirc\circ\circ\redcirc\redcirc\longrightarrow\bluecirc\circ\circ\bluecirc\circ\redcirc\redcirc$.
Notice that the final part, $\cdots\bluecirc\circ\redcirc\redcirc$, the hole can no longer propagate without violating domain-wall and charge number, therefore, to ensure the symmetries are conserved we need to remove a single hole.
The sub-sector is then formed by $\nb=\nr=2$ with $\nh=2$, where the color ordered string is $\bluecirc\bluecirc\redcirc\redcirc$.
Consequently, these immobile particles are inert in otherwise nearest-neighbor hopping.
Therefore, each spin sequence after the map has immobile particles and boundaries that begin(end) with $\redcirc(\bluecirc)$ removed.
After these degrees of freedom are taken into account the final configuration is composed of, $n_{\redcirc}$, $n_{\bluecirc}$ and mobile holes $n_{\circ}$ that label a fragmented sub-sector.

\section{Entanglement Entropy for Exact Eigenstate}
\label{appsec:entanglement}
In Eq.(5) of the main text, we construct an exact eigenstate for each invariant sub-sector of the Hamiltonian Eq.(3).
To calculate the half-chain entanglement entropy of this state, we first perform a Schmidt-decomposition with the cut placed in the center of the chain
\bea
\ket{\Psi}=\sum\limits_i \lambda_i \ket{\Psi^L_i} \ket{\Psi^R_i},
\eea
where $\ket{\Psi^{R/L}_i}$ is a set of orthonormal states on the left(right) half of the chain and the $\lambda_i$ is the Schmidt coefficient.
Due to the normalization of $\ket{\Psi}$, the Schmidt coefficients obey $\sum \lambda^2_i=1$.
From $\lambda_i$, the R\'enyi entropy of index $\alpha$ can be calculated as
\bea
S^{(\alpha)} = \frac{1}{1-\alpha} \log \sum\limits_i \lambda_i^{2\alpha}. 
\eea
Because of the wavefunctions factorable structure and the conserved number of holes, the orthornormal basis, $\ket{\Psi^{L(R)}}$, can be labelled by the number of holes on each side.
Therefore the Schmidt decomposition is re-written as
\bea
\ket{\Psi}=\sum\limits_{m} \lambda_m \ket{\Psi^L_m} \ket{\Psi^R_{\nh-m}}
\label{appeq:schmdit}
\eea
where $m$ is the hole number on the left, and $\nh-m$ is the hole number on the right. Note that $m$ can be both integer and half-integer, the half-integers correspond to the case where the entanglement cut is through a hole which takes two sites in the spin representation.
The number of holes on the left(right) side is between $0$ and $N/4$, imposing a constraint on the range of $m$,
\bea
\begin{cases}
0\leq m \leq \nh, & \nh\leq \frac{N}{4} \\ 
\nh-\frac{N}{4} \leq m \leq \frac{N}{4}, & \nh>\frac{N}{4}
\end{cases}
\eea
As a result, there are at most $\min (2\nh+1, N-2\nh+1)$ nonzero singular values.
The entanglement entropy is therefore upper bounded by $S \leq \log( \min (2\nh+1, N-2\nh+1)$.
However, when either $\nr+\nb$ or $\nh$ stays finite as $N\rightarrow\infty$, the exact state has area-law entanglement entropy.
When $\nh/N$ stays fixed as $N\rightarrow\infty$, the entanglement scales as $\log(N)$ or remains area law depending on values of $q$, which is elaborated on below.
In any case, however, the state has sub-volume law entanglement, violating ETH. 

In order to obtain the entanglement scaling for the case when $\nh$ increases linearly with $N$, we look into the Schmidt decomposition Eq.~(\ref{appeq:schmdit}) more closely.
The states $\ket{\Psi^L_m}$ are,
\bea
\ket{\Psi^L_m} = \frac{1}{\mathcal{N}_m(q)}\exp (-\log (q) \hat{\mathcal{P}}_L/2)\sum\limits_m \ket{\psi^L_m}.
\eea
The summation is over all states on the left partition containing $m$ holes, and $\mathcal{N}_m(q)$ is the normalization factor.
The state $\ket{\Psi^R}_m$ is similarly constructed,
\bea
\ket{\Psi^R_m} = \frac{1}{\mathcal{N}_m(q)}\exp (-\log (q) \hat{\mathcal{P}}_R/2)\sum\limits_m \ket{\psi^R_m}.
\eea
The Schmidt coefficients are
\bea
\lambda_m (q) =\frac{\mathcal{N}_m(q)\mathcal{N}_{\nh-m}(q)}{\mathcal{N}(q)}.
\eea
We now provide the explicit derivation of the R\'enyi entropy for the exact state with the sub-sector labeled by $\bredcirc\redcirc\cdots\redcirc\bluecirc\cdots\bluecirc\bbluecirc$ with $\nr = \nb$ in the large $N$ limit.
Due to the fixed boundaries, $\nr+\nh=N/2-1$. 
As we will show below, the large $N$ scaling of the R\'enyi entropy depends on whether $q=1$, $q<1$ or $q>1$. 

\subsection{Case of \texorpdfstring{$q=1$}{Lg}}
When $q=1$ the exact eigenstate is an equal weight superposition independent of the dipole moment. 
As a result the normalization factor, $\mathcal{N}(q)^{2} = \text{tr}(q^{-\hat{\mathcal{P}}}) = \dim$ which can be found using combinatorial methods.
We find the total Hilbert space dimension and the dimension of each half with fixed hole number to be
\bea
\mathcal{N}^2 = 
\begin{pmatrix}
2\nr+\nh \\
2\nr
\end{pmatrix}, \quad 
\mathcal{N}^2_m = 
\begin{pmatrix}
 \lfloor -m \rfloor + \nh+\nr \\
 \lfloor m \rfloor
\end{pmatrix} 
\eea
where $\lfloor \cdots \rfloor$  is the floor function. When $m$ is a half-integer, it implies that a hole ($\downarrow\uparrow$) is in the middle of chain. 
The Schmidt coefficients $\lambda$ can be computed straightforwardly
\bea
\lambda_m^2  = 
\begin{pmatrix}
 \lfloor -m \rfloor + \nh+\nr \\
 \lfloor m \rfloor
\end{pmatrix} 
\begin{pmatrix}
 \lfloor m  \rfloor  -\nh \\
 \lfloor \nh - m  \rfloor
\end{pmatrix} \bigg/
\begin{pmatrix}
2\nr+\nh \\
2\nr
\end{pmatrix}
\eea
Then the R\'enyi entropy before taking the $\log$ is
\bea
\exp &((1-\alpha)S^{(\alpha)}) =\sum\limits_m \lambda_m^{2\alpha}
=\sum\limits_m 
\begin{pmatrix}
 \nr+\nh + \lfloor -m \rfloor  \\
\lfloor m \rfloor
\end{pmatrix} ^\alpha
\begin{pmatrix}
 \lfloor m \rfloor + \nr \\
 \nh+ \lfloor -m \rfloor
\end{pmatrix}^\alpha \bigg/
\begin{pmatrix}
2\nr+\nh \\
2\nr
\end{pmatrix}^\alpha\\
\eea
The R\'enyi entropy can be calculated in the thermodynamic limit, $N\rightarrow \infty$.
We further set $\alpha=2$ and obtain
\bea
S^{(2)}=
\begin{cases}
   2\nh\log 2 -\log \begin{pmatrix}2\nh\\\nh\end{pmatrix}  & \text{if $N \rightarrow \infty$ and $n_{\circ} \ll N$} \\
   \\
   4\nr\log 2 -\log \begin{pmatrix}4\nr\\2\nr\end{pmatrix}  & \text{if $N \rightarrow \infty$ and $\nr \ll N$} \\
   \\
  \frac{1}{2}\log(N) + \frac{1}{2}\log(\pi(1-\gamma)(2-\gamma)\gamma)-\log ((1-\gamma)^2+1)  & \text{if $N \rightarrow \infty$ and $\gamma =n_{\circ}/N$  finite}
  \end{cases}
\eea
As expected, $S$ exhibits area law when either $\nh$ or $\nr$ stay fixed as $N\rightarrow\infty$ and scales logarithmically with $N$ if the ratio $\nh/N$ remains fixed instead.
In the following sections we determine the behavior away from the critical point.

\subsection{Case of \texorpdfstring{$q>1$}{Lg}}
We now consider the R\'enyi entropy when the deformation $q>1$, where we are interested in the regime with $\gamma = n_{\circ}/N$ fixed as $N\rightarrow\infty$.
In this case, the exact eigenstate is a superposition of all states in the computational basis weighted by the its dipole moment.
The largest contribution is from the state of largest dipole moment, given by
\bea
\ket{\bredcirc\redcirc\cdots\redcirc\overbrace{\circ\cdots\circ}^{\nh}\bluecirc\cdots\bluecirc\bbluecirc}.
\label{appeq:large_dip}
\eea
The states with $\gamma N -\mathcal{O}(1)$ holes between the two color domains have the same dipole moment density as Eq.~(\ref{appeq:large_dip}), and contribute to the exact state.
Similarly, other states have a small density, $|\hat{\mathcal{P}}|$ and are exponentially suppressed by large $N$.
The contributing states have identically half the number of holes on each half of the chain.
As a result, the Schmidt coefficient $\lambda_m$ is 1 when $m=\nh/2$ and 0 otherwise.
This suggests that for $q>1$ and $N\rightarrow\infty$ the entanglement entropy for the exact state vanishes.
When $N$ is finite the entanglement decays to zero rapidly as $q$ increases from 1.

We confirm this behavior numerically using MPS for $N=100$ to $400$ in Fig.2(b).
This behavior can also be seen in Fig.~\ref{fig:diff_ent}(d) where the entanglement entropy for the exact state(red point) is approximately zero.
An interesting direction would be to study the critical regime, $(q-1)\sim 1/N$.

\subsection{Case of \texorpdfstring{$q<1$}{Lg}}
In this case the entanglement entropy has area or logarithmic scaling depending on the ratio, $\gamma = n_{\circ}/N$.
We consider the same sub-sector with $\nr=\nb$ with $q<1$ and finite in the large $N$ limit.
Here the superposition will be dominated by a uniform superposition of states minimizing the dipole moment amplitude,
\bea\label{appeq:small_dip}
\frac{1}{\nh+1}\sum\limits_{m=0}^{n_\circ}\ket{\bredcirc\overbrace{\circ \cdots \circ}^m \redcirc\cdots\redcirc\bluecirc\cdots\bluecirc\overbrace{\circ\cdots\circ}^{\nh-m} \bbluecirc}
\eea
for $0\leq m \leq \nh$.
Unlike the previous case, there is no unique state but a $(\nh+1)$ dimensional manifold of states.
Now let us consider the half-chain entanglement entropy for the uniform superposition of these states in the spin representation.
When $\nr \geq \nh$ or equivalently $\gamma\leq 1/4$, the particle domain $\redcirc\cdots \redcirc \bluecirc \cdots\bluecirc$ takes more than half of the chain.
As a result, the half-chain cut intersects the particle domain for each of the $(\nh+1)$ configurations.
Therefore the states $\Psi^{R/L}$ from the Schmidt decomposition are
\bea
\ket{\Psi^L_m} = \ket{\bredcirc\overbrace{\circ\cdots\circ}^m\redcirc\cdots}, \quad
\ket{\Psi^R_{\nh-m}} = \ket{\cdots\bluecirc\overbrace{\circ\cdots\circ}^{\nh-m}\bbluecirc}
\label{appeq:small_dip_schmdit}
\eea
where $m$ range from $0$ to $\nh$.
Therefore, the state Eq.~(\ref{appeq:small_dip}) in this case has a flat entanglement spectrum with $(\nh+1)$ Schmidt coefficients.
The half chain entanglement entropy is just $\log (\nh+1)$, which scales as, $\gamma \log (N)$.

When $\nr < \nh$ or $\gamma>1/4$, the cut is within the hole domain for some of the configurations, for which the entire particle domain is within the left side or the right side.
Therefore, one half of the chain is frozen and the corresponding Schmidt decomposition becomes,
\bea
&\ket{\Psi^L_{(\nh+\nr)/2}} = \ket{\bredcirc\overbrace{\circ\cdots\circ}^{(\nh+\nr)/2}}, \quad \ket{\Psi^R_{(\nh-\nr)/2}} = \frac{1}{(\nh-\nr)/2+1}\sum\limits_{m=0}^{(\nh-\nr)/2}\ket{\overbrace{\circ \cdots \circ}^m \redcirc\cdots\redcirc\bluecirc\cdots\bluecirc\overbrace{\circ\cdots\circ}^{\nh-m} \bbluecirc}\\
&\ket{\Psi_{(\nh+\nr)/2}^R} = \ket{\overbrace{\circ\cdots\circ}^{(\nh+\nr)/2)}\bbluecirc},\quad \ket{\Psi^L_{(\nh-\nr)/2}} = \frac{1}{(\nh-\nr)/2+1}\sum\limits_{m=0}^{(\nh-\nr)/2}\ket{\bredcirc\overbrace{\circ \cdots \circ}^m \redcirc\cdots\redcirc\bluecirc\cdots\bluecirc\overbrace{\circ\cdots\circ}^{\nh-m} }
\eea
The Schmidt coefficients associated with the two pairs of state are 
\bea
\quad \lambda^2=\frac{(\nh-\nr)/2+1}{n_\circ+1}
\eea
We have assumed $\nh+\nr$ is an even number for simplicity
In addition, there are another $\nr-1$ pairs of states similar to those in Eq.~(\ref{appeq:small_dip_schmdit}) with the same Schmidt coefficients $\lambda^2=1/(\nh+1)$.
From these Schmidt coefficients, we obtain the R\'enyi entropy for $\gamma>1/4$ in the thermodynamic limit as 
\bea
S = 2\log (4\gamma) -2\log(4\gamma-1) -\log 2
\eea
which clearly has no dependence on the chain length demonstrating area law behavior.

\section{The Matrix Product Representation of the Exact Eigenstate}
\label{appsec:MPS_scar}
In order to study the entanglement structure of the exact eigenstate we construct a matrix product ansatz in the spin representation.
Each sub-sector hosts an exact eigenstate with zero energy density which is a weighted superposition of each basis state and therefore sits in the middle of the spectrum and has the form
\bea
\ket{\Psi(q)} = \frac{1}{\mathcal{N}}\mathbf{P}\left( q^{-\hat{\mathcal{P}}/2} \sum\limits_{i=1}^{2^N} \ket{\psi}_i\right)
\label{appeq:exact}
\eea
where $\mathbf{P}$ projects the state into the invariant sub-sector designated by the color sequence with the normalization factor, $\mathcal{N}$.
Our task is to construct an exact MPS for the state
\bea
 \ket{\Psi(q)}=\sum\limits_{\{s\}} A^{s_1}_1 A^{s_1}_2 \cdots A^{s_1}_N  \ket{s_1s_2\cdots s_N}
\eea
here the matrices $A^{s}_i$ are to be determined.
The summation in Eq.~(\ref{appeq:exact}) is over the entire Hilbert space, which allows the state before the projection to be decomposed as a product state
\bea
 q^{-\hat{\mathcal{P}}/2} \sum\limits_{i=1}^{2^N} \ket{\psi}_i=\prod\limits_{i=1}^N \left( q^{-i/2}\ket{\uparrow}_i + q^{i/2}\ket{\downarrow}_i\right ).
\label{appeq:product_state}
\eea
Therefore, the difficulty of constructing the MPS depends on the correct matrix product operator for the projection, $\mathbf{P}$. Once obtained the exact eigenstate follows.
The structure of the projection operator depends on the sub-sector defined through the color sequence.
Generically, the projection operator is defined as
\bea
\mathbf{P} = \sum_{i}^{\text{dim}} \ket{\psi_i}\bra{\psi_i}
\eea
where the sum is now over the states in the fragmented sub-sector. Using standard techniques we seek the projection as a matrix product operator(MPO) of the form
\bea
\mathbf{P} =V_L^T\mathbb{P}_1 \mathbb{P}_2 \cdots \mathbb{P}_N V_R
\eea
where $\mathbb{P}_i$ is a $d\times d $ matrix of local operators living on site $i$, where $d$ is the bond dimension of the MPO.
$V_L$ and $V_R$ are $d$-dimensional boundary vectors to ensure that the result is an operator.

The MPO can be constructed by induction: 
we require that
\bea 
V_L^T\prod\limits_1^r \mathbb{P}_i = (\mathbf{P}_r(0), \mathbf{P}_r(1/2), 
\mathbf{P}_r(1), \cdots \mathbf{P}_r(\nh)
\eea
where $\mathbf{P}_r(n)$ is the projection operator acting on the first $r$ sites to ensure that the states contain $n$ holes after mapping.
The projection operator, $\mathbf{P}_{r}(n)$, can be obtained by combining $\mathbf{P}_{r-1}(n)$, $\mathbf{P}_{r-1}(n-1/2)$ and the operators defined on site $r$.
The combination, which depends on the color sequence, specifies the matrix operator, $\mathbb{P}_{r}$.
To be concrete, consider the sub-sector given by $\bredcirc\redcirc\cdots\redcirc\bluecirc\cdots\bluecirc\bbluecirc$. We have the following iteration rules 
\bea
\mathbf{P}_{r}&(n+1/2)  = \mathbf{P}_{r-1}(n) P^\downarrow_{r}\\
\mathbf{P}_{r}&(n) =
\begin{cases}
\mathbf{P}_{r-1}(n-1/2)P^\uparrow_{r} + \mathbf{P}_{r-1}(n) P^{\sigma(r-2n) }_{r}, &2\nh-2n<N-r\\
\mathbf{P}_{r-1}(n-1/2)P^\uparrow_{r}, &2\nh-2n\geq N-r
\end{cases}
\eea
The site index $r$ goes from $2$ to $N-1$. When $r=1$, $\mathbf{P}_1(n)=P^\uparrow_1\delta_{n,0}$ and the first spin is fixed to be up.
When $r=N$, $\mathbf{P}_{N}(\nh)=\mathbf{P}_{N-1}(\nh)P^\downarrow_{N}$ and the last spin is fixed to be down. From the iteration rules, one can construct the MPO accordingly. The naive maximal bond dimension of the MPO constructed using this approach is $2\nh+1$. However, at a given site, not all $2\nh+1$ projectors $\mathbf{P}_r(n)$ are physical because the constraint that the $n$ holes has to fit into $r$ sites and the remaining $\nh-n$ holes has to fit into the remaining $N-r-1$ sites. Following the iteration rules given above, the non physical projectors becomes 0, leading to reduction of the bond dimension. One can show that the maximal bond dimension in this case is $\min(2\nh+1, 2\nr)$. 
The MPO of the projection operator for other sub-sectors can be constructed in a similar manner.
With the MPO of the projection operator, one can immediately obtain the MPS of the exact eigenstate by acting on the product state Eq.~(\ref{appeq:product_state}).

\section{MPO of the Hamiltonian and Details on Simulating the Dynamics}
\label{appsec:TDVP}
In this section, we write down the MPO of the Hamiltonian in Eq.(3), which is used in standard TDVP algorithm to evolve matrix product states. The Hamiltonian is
\bea
H = \frac{4}{q}\sum\limits_{i=1}^{N-2} P_{i-1}^\uparrow \ket{\Phi(q)}\bra{\Phi(q)}_{i,i+1} - \ket{\Phi(q)}\bra{\Phi(q)}_{i,i+1} P_{i+2}^\downarrow
\eea
where the un-normalized state is defined as, $\ket{\Phi(q)}_{i,i+1} = \ket{\uparrow}_i\ket{\downarrow}_{i+1}-q\ket{\downarrow}_i\ket{\uparrow}_{i+1}$.
It is convenient to re-express the two site projectors in terms of Pauli operators 
\bea
 \frac{4}{q} \ket{\Phi(q)}\bra{\Phi(q)}_{i,i+1} =\frac{4}{q}P^\uparrow_i P^\downarrow_{i+1} + 4q P^\downarrow_i P^\uparrow_{i+1} - 2\left( \sigma^x_i \sigma^x_{i+1}+\sigma^y_i \sigma^y_{i+1}\right).
\eea
We would like to convert the Hamiltonian to a matrix product operator
\bea
H  = V_L^T \mathbb{H}_1\mathbb{H}_2\cdots \mathbb{H}_N V_R
\eea
where $\mathbb{H}_i$ is a matrix of operators defined on each site and $V_{L/R}$ are boundary vectors.
The MPO is used as the input for the tensor network simulation of the dynamics.  
Following the standard approach, one can show that
\bea
V_L &= (1, 0, 0, 0, 0, 0)^T, \quad V_R =(0, 0, 0, 0, 0, 1)^T,  \quad
\mathbb{H}_i &=
\begin{pmatrix}
I_i  & P^\uparrow_i & 0 & \vec{O}_{L,i}^T & 0 & 0\\
0  & 0 &\vec{O}_{L,i}^T & 0 & 0 & 0\\
0  & 0 & 0 & 0 & 0 & \vec{O}_{R,i} \\
0  & 0 & 0 & 0 & \vec{O}_{R,i} & 0 \\
0  & 0 & 0 & 0 &  0   & -P^\downarrow_i \\
0  & 0 & 0 & 0 &  0   & I_i \\
\end{pmatrix}
\eea
where 
\bea
\vec{O}_{L,i} &= \left( \frac{4}{q}P^\uparrow_i,4q P^\downarrow_i, -4 \sigma^x_i, -4 \sigma^y_i \right)^T, \quad \vec{O}_{R,i} &= \left( P^\downarrow_i, P^\uparrow_i, \sigma^x_i,  \sigma^y_i \right)^T
\eea
so that $\vec{O}_{L,i}^T \vec{O}_{R,i} =\frac{4}{q} \ket{\Phi(q)}\bra{\Phi(q)}_{i,i+1}$. The bond dimension of the MPO is 12.  One can check that 
\bea
V_L^T \prod\limits_i^N \mathbb{H}_i = \left(\mathbbm{1}, P^\uparrow_{N}, P^\uparrow_{N-1}\vec{O}^T_{L,N}, \vec{O}^T_{L,N}, O_{N-1,N}, H\right)
\eea
which multiplied by the right boundary vector gives the Hamiltonian. This completes the construction of the MPO for the Hamiltonian and we now discuss how unitary evolution using the time-dependent variational algorithm(TDVP)~\cite{Haegeman,cirac}.

The TDVP-MPS algorithm has the advantage of conserving the state's energy during time evolution explicitly and therefore is particularly suitable for simulating dynamics governed by a time-independent Hamiltonian.
However, the TDVP does not respect the emergent conserved color sequence. 
Consequently, the algorithm can introduce a state that does not belong to the fragmented sub-sector introducing error. To avoid this, we combine the Hamiltonian MPO with the projection operator discussed in Sec.~\ref{appsec:MPS_scar} in the TDVP simulation.  

\section{Dynamics and Eigenstate Properties from Exact Diagonalization}
\label{appsec:dynamics}
\begin{figure}
\includegraphics[width=0.90\textwidth]{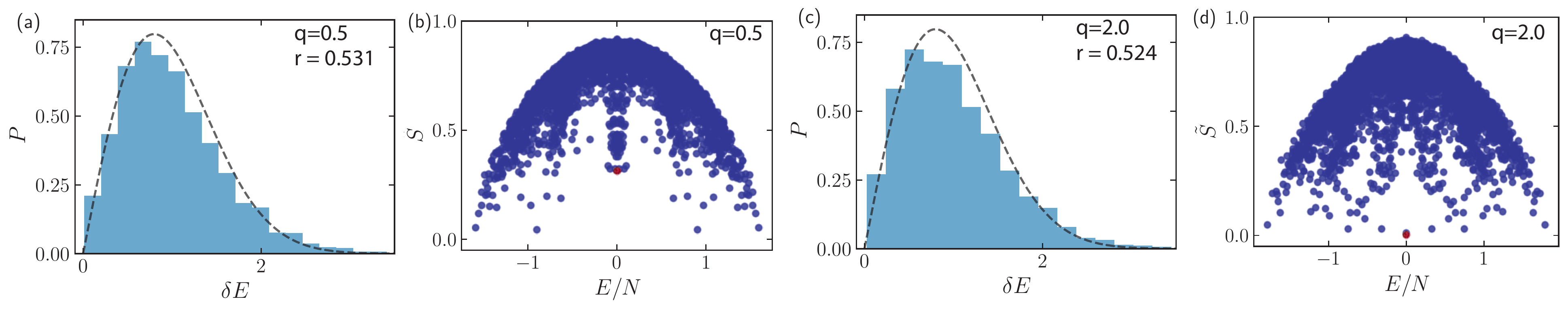}
\caption{(a) and (c) Level spacing between adjacent eigenvalues for the color sequence, $\redcirc\redcirc\redcirc\redcirc\bluecirc\bluecirc\bluecirc\bluecirc$, with $\nh=10$ and $\dim=8008$.
The inset gives the deformation value and average level ratio parameter, $\braket{r}$.
(b) and (d) Normalized half-chain entanglement entropy, $\tilde{S}$, for the same sub-sector using exact diagonalization.
For $q=0.5$ a majority of the eigenstates form an ETH band but the presence of outlying states remain visible.
In the regime, $q=2.0$, the ETH band begins to collapse and further scarring is observed.
The red point is the entanglement for the exact eigenstate, Eq.(5) with $\tilde{S}\sim0$ for $q=2.0$.
}%
\label{fig:diff_ent}
\end{figure}
This section studies the eigenstate entanglement entropy and dynamics through exact diagonalization for a pair of deformation values, $(q, 1/q)$.
We first checked that over the range, $q \in [0.5, 2.0]$, the level statistics remain Wigner-Dyson, illustrated in Fig.~\ref{fig:diff_ent}(a) and (c).
In Fig.~\ref{fig:diff_ent}(b) and (d) we determine the eigenstate bipartite entanglement entropy in the sub-sector defined by, $\bredcirc\redcirc\redcirc\redcirc\bluecirc\bluecirc\bluecirc\bbluecirc$ with $\nh=10$.
For $q<1.0$, there is a distinct ETH band; however, the quantum many-body scars persist.
At $q>1$, the entanglement distribution begins to collapse and broaden, but weakly entangled states far below the band are apparent.
\begin{figure}[t]
\includegraphics[width=\textwidth]{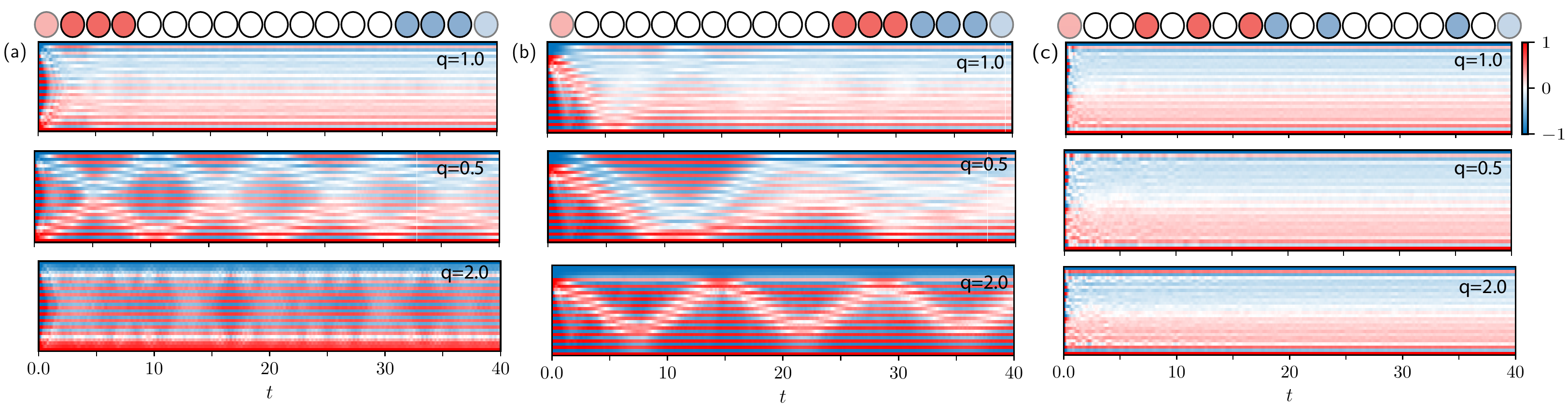}
\caption{(a-c) Dynamics of $\langle \sigma_{i}^{z}(t) \rangle$ from exact diagonalization for $q=1.0, 0.5, 2.0$ prepared in the state with (a) large dipole amplitude (b) small dipole amplitude and (c) a random state.
The non-ergodic dynamics are apparent in (a) and (b) at $q=0.5(q=2.0)$ where coherent oscillations are present corresponding to particle scattering.
While at $q=2.0$ in (a) the dynamics are nearly static since the prepared state is approximately an eigenstate.
(c) Demonstrates the quench dynamics with a random state, for which thermalization occurs across all deformation values.}
\label{fig:dyn}
\end{figure}

To accompany the TDVP-MPS simulations, we perform exact diagonalization in the spin representation within the same sub-sector and compute the operator's expectation time-evolution, $\braket{\sigma^z_i(t)}$.
Illustrated in Fig.~\ref{fig:dyn} we quench the system upon being prepared in the large(small) dipole state.
Since the domains are small, the timescale over which the boundary term becomes important is short.
As explained in the main text, due to the boundary term, the energy difference between the two configurations $\bredcirc\redcirc\cdots\redcirc\cdots$ and $\bredcirc\circ\redcirc\cdots\redcirc\cdots$ is $V(q)-h(q)=4q$ instead of $V(q)=2(1/q+q)$. As a result, the boundary term reduces the energy cost of detaching the domain from the boundary for $q<1$, causing ballistic domain propagation.
The collisions of the two domains contribute to the entanglement oscillations in Fig.2(d), where the period is equivalent to the timescale between the domain scattering, Fig.~\ref{fig:dyn}(a).
For the large dipole state at $q=2.0$, the dynamics are nearly stationary because the boundary term increases the energy cost to move the domain. Recall that Hamiltonian at $q=0.5$ and $q=2.0$ only differ by the boundary term.
For the small dipole state shown in Fig.~\ref{fig:dyn}, the red domain oscillates for both $q=0.5$ and $q=2.0$. This follows because when the red domain detached the blue domain, $\circ\cdots\circ\redcirc\cdots\redcirc\circ\bluecirc\cdots\bluecirc$, the nearest-neighbor energy from the $\redcirc\circ$ and $\circ\bluecirc$ contribute an energy $V(q)$ and $-V(q)$ which cancel, ensuring the net energy at zero.
Due to this cancellation, the holes between the two domains act as an effective boundary by maintaining the red domain's energy to propagate collectively. On the other hand, the dynamics of the blue domain are affected by the boundary term. 
We also check that random product states in the fractured sub-sector obey Kyrlov-ETH and thermalize in a short time scale, Fig.~\ref{fig:dyn}(c).

\begin{figure}[h]
\includegraphics[width=0.6\textwidth]{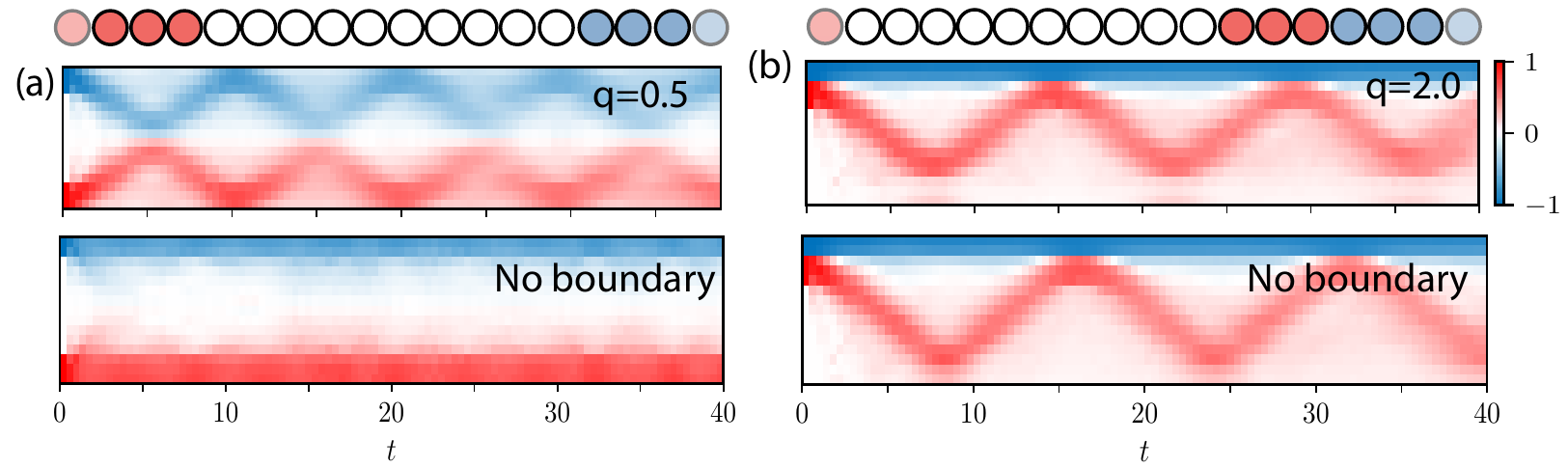}
\caption{(a-b) Dynamics of $\langle n_{i,\redcirc}(t)\rangle-\langle n_{i,\bluecirc}(t)\rangle$ using exact diagonalization of the effective Hamiltonian, Eq.(4).
The parameters are $N=28$ with the color sequence $\bredcirc\redcirc\redcirc\redcirc\bluecirc\bluecirc\bluecirc\bbluecirc$ and $\nh=10$ for $t=40$ with a time step, $dt=0.05$.
The bottom panels are the effective Hamiltonian with the boundary term removed.
By quenching the system prepared in the state with the smallest dipole amplitude without the boundary, non-ergodic behavior persists.
This behavior follows from the nearest-neighbor energy canceling between the red(blue) domain; thus, the holes induce an effective boundary allowing the red domain to propagate collectively.
}
\label{fig:effective_dynamics}
\end{figure}

In Fig.~\ref{fig:effective_dynamics} we simulate the effective Hamiltonian, Eq.(4) using exact diagonalization with and without the boundary term to explicitly demonstrate its effect.
We focus on two cases that exhibit long time coherent oscillation: (a) $\ket{\bredcirc\redcirc\cdots\redcirc\circ\cdots\circ\bluecirc\cdots\bluecirc\bbluecirc}$ at $q=0.5$, (b) $\ket{\bredcirc\circ\cdots\circ\redcirc\cdots\redcirc\bluecirc\cdots\bluecirc\bbluecirc}$ at $q=2.0$. 
In the first case shown in Fig.~\ref{fig:effective_dynamics}(a), when there is no boundary term, the dynamics are nearly static due to the boundary term no longer minimizing the energy cost to move the domains.
A different behavior arises in Fig.~\ref{fig:effective_dynamics}(b) where coherent oscillations persist in the absence of the boundary.

\section{Floquet Unitary Circuit}
\label{appsec:circuit}
\begin{figure}[h]
\includegraphics[width=0.9\textwidth]{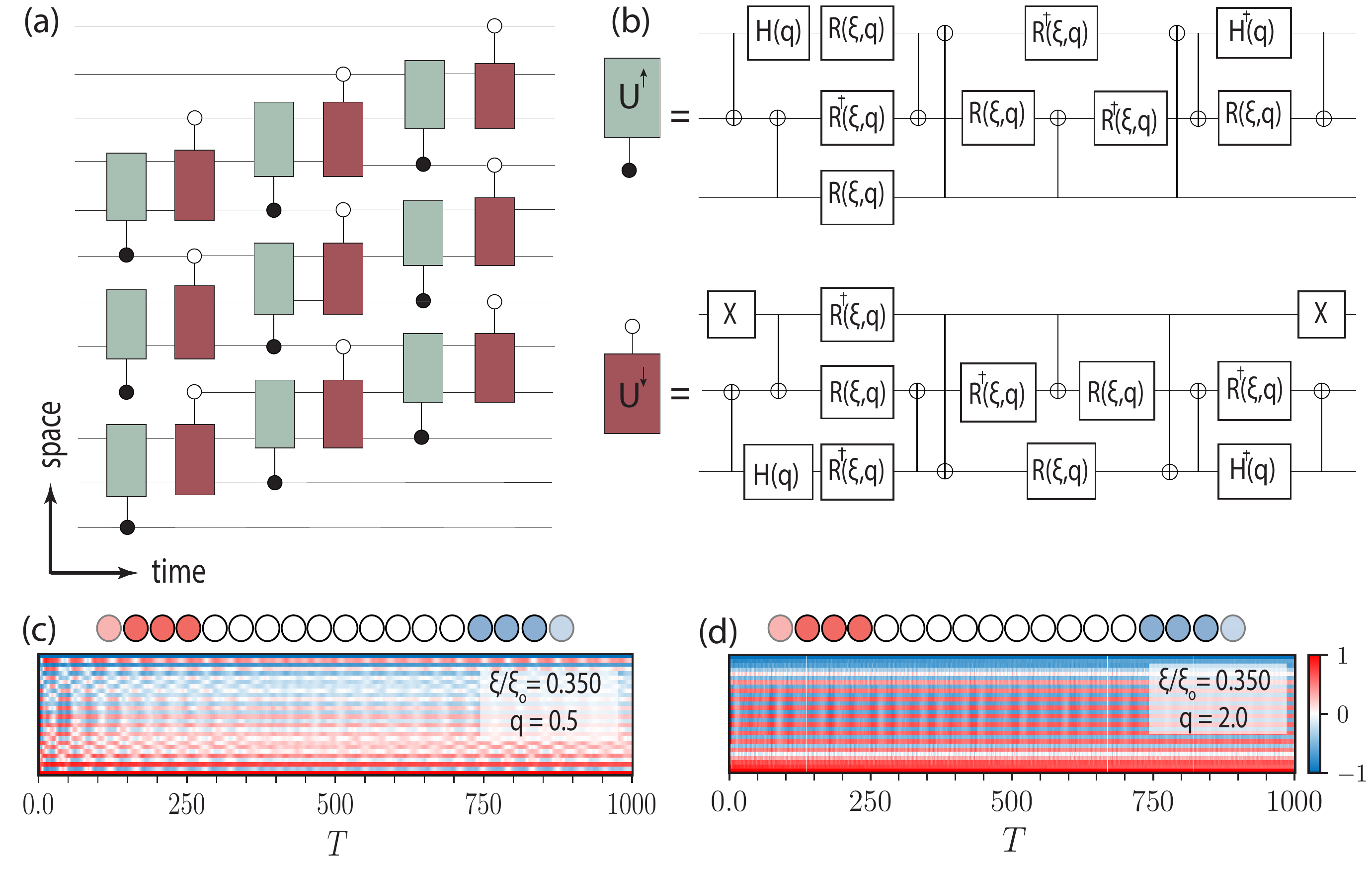}
\caption{(a) Quantim circuit diagram of the Floquet operator consisting of six layers each containing three-site gates.
The circles represent the control-bit with the filled getting projected on by, $\ket{\uparrow}\bra{\uparrow}$ and the empty by $\ket{\downarrow}\bra{\downarrow}$.
(b) Efficient gate decomposition into CNOT and on-site gates.(c-d) Quench dynamics of the Floquet operator from the state with the large dipole amplitude in the sub-sector generated by $\nr=\nb=3$ and $\nh=10$ for 1000 periods. The circuit is evaluated at $(\xi /\xi_{o}=0.350, q=0.5)$ in (c) and $(\xi /\xi_{o}=0.350, q=2.0)$ in (d). Note that the stationary behavior still persists for $q>1.0$, while for $q<1.0$ there are coherent oscillations indicating the non-ergodic dynamics remain present.}
\label{fig:circuit_decomp}
\end{figure}

Here we provide details for an efficient gate decomposition for the Floquet circuit for arbitrary $\xi$ and $q$ given in Eq.(6), shown in Fig.~\ref{fig:circuit_decomp}.
The three-site gate is decomposed into a series of CNOT gate and single-site gate $H(q)$ and $R(q,\xi)$ where
\bea
H(q) =\frac{1}{\sqrt{1+q^2}} (qI+i \sigma^y),\quad
R(\xi,q) = \exp\left (-i \frac{\xi (1+q^2)}{2q} \sigma^z.
\right)=\exp\left(-i \frac{\pi}{4}\frac{\xi}{\xi_o(q)}\sigma^z \right)
\eea
When $q=1$, $H(q)$ becomes the Hadamard gate, and when $\xi=\xi_o/2$, $R(q,\xi)$ becomes the T gate.
At $\xi=\pi/8$ and $q=1$, the three-site gate becomes the Fredkin gate.
Recent advancements in quantum dots and photonic circuits has made experimental realization of this gate attainable~\cite{kim2018efficient, ono2017implementation, kang2020optical,gambetta2020long,feng2020quantum,petiziol2020quantum}.
Through these experimental methods, the discussed fragmentation structure and resulting non-ergodic dynamics are feasible in the imminent future utilizing near-term intermediate-scale quantum devices.
\bibliography{references}